\documentclass[aps,prb,reprint,twocolumn,floatfix,superscriptaddress,longbibliography]{revtex4-2}

\usepackage[T1]{fontenc}
\usepackage[utf8]{inputenc}

\usepackage{amsmath,amssymb,bm}
\usepackage{physics}

\usepackage{graphicx}
\usepackage[colorlinks,citecolor=blue,linkcolor=blue,urlcolor=blue]{hyperref}

\begin{document}
\title{Comment on\\
``Electrostatics-induced breakdown of the integer quantum Hall effect in cavity QED''}
\author{C.~Ciuti}
\affiliation{Universit\'{e} Paris Cit\'{e}, CNRS, Mat\'{e}riaux et Ph\'{e}nom\`{e}nes Quantiques, 75013 Paris, France}
\author{G.~Scalari}
\author{J.~Faist}
\affiliation{Institute for Quantum Electronics, ETH Zurich, CH-8093 Zurich, Switzerland}
\affiliation{Quantum Center, ETH Zürich, 8093 Zürich, Switzerland}

\maketitle

In their theoretical manuscript, Andolina \textit{et al.}~\cite{Andolina2025}
propose an electrostatic mechanism for the breakdown of integer quantum Hall
plateaus observed in GaAs two-dimensional electron gases coupled to cavities based
on metallic split-ring resonators~\cite{Appugliese2022}. Their model assumes that
metallic elements of a split-ring resonator placed in  proximity to the
semiconductor induce, via an image-charge potential, a net attractive electrostatic
potential for electrons, generating a local energy minimum near the mesa edge. This
putative potential pocket is argued to support two counterpropagating edge channels
and to enable sufficient backscattering to affect Hall quantization. In practice,
Ref.~\cite{Andolina2025} does not calculate  the transverse Hall resistance and
addresses a two-terminal configuration.

We disagree with the central claims of Ref.~\cite{Andolina2025} for two reasons:
(i) the electrostatic treatment of the semiconductor interface is incorrect, and
(ii) the resulting predictions are incompatible with experimental observations.

The treatment assumed in Ref.~\cite{Andolina2025} contradicts the
well-established electrostatics of GaAs/AlGaAs heterostructures
\cite{SzeNg2007,Heinzel_Mesoscopic_Wiley, Glazman1992,Geier2020,Glazman1992}. In a semiconductor
two-dimensional electron gas, electrostatics is governed by the negative charge density of the 2D electron gas,
together with the positive background charge of ionized donors. The electrostatic
potential follows from Poisson’s equation, which depends on this total charge
density and is supplemented by appropriate boundary conditions. For an etched GaAs
mesa edge, the boundary condition is imposed by Fermi-level pinning at the
GaAs/air interface \cite{Spicer1979, Heinzel_Mesoscopic_Wiley,WeberPasquarello2014}, caused by the very high density of mid-gap surface states. This
pinning produces a depletion region and a smooth, flattened parabolic confinement
near the mesa edge. As shown explicitly in the
Appendix, even in the presence of a nearby metal, when the correct macroscopic
charge densities and boundary conditions are taken into account, the lateral
confinement is strictly monotonic and does not develop the local minimum postulated
in Ref.~\cite{Andolina2025}. Andolina {\it et al.} ~\cite{Andolina2025} effectively treats the
problem as that of a single electron interacting with a nearby metal via an
image-charge potential, while neglecting the collective electrostatics of the
two-dimensional electron gas, the donor background, and the surface boundary
conditions. In other words, Ref.~\cite{Andolina2025} considers interactions with
image charges while disregarding the direct interaction with the real
charges.

Notice that if such a potential pocket existed, one could routinely create
one-dimensional channels simply by depositing metal on etched mesa boundaries,
which is never observed in GaAs devices. Likewise, quantum point contacts in the
presence of a magnetic field~\cite{Geier2020} would not function as they do
experimentally, since the putative pocket would disrupt the edge channels on which
their operation relies.

In addition to being incompatible with the standard electrostatics of GaAs
surfaces, the transport behavior predicted in Ref.~\cite{Andolina2025} is at odds
with the experimental observations of Appugliese \textit{et al.}~\cite{Appugliese2022}.
The model of Ref.~\cite{Andolina2025} addresses a $2~\mu\mathrm{m}$-wide device and
computes only the two-terminal resistance, whereas the experiment measures longitudinal and Hall resistance
 in $40~\mu\mathrm{m}$-wide Hall bars. All cavity devices displayed similar
collective vacuum Rabi frequencies and quantitatively comparable modifications of
quantum Hall features, with the corresponding quantitative figure of merit $\rho_{xx}^\nu$ changing by only about a factor of three.
By contrast, the mechanism proposed in Ref.~\cite{Andolina2025} would imply that the
effect for samples with a $20~\mu\mathrm{m}$ cavity separation should be totally
negligible compared with that for a $250~\mathrm{nm}$ separation. Concerning the
role of cavity-mediated hopping~\cite{Ciuti2021,Arwas2023,Borici2025}, Andolina \textit{et al.}
\cite{Andolina2025} neither compute the corresponding quantum Hall transport
observables nor analyze the scaling with the system size.

In summary, the treatment proposed in Ref.~\cite{Andolina2025} is incompatible with
the electrostatics of GaAs heterostructures and with the functioning of
standard GaAs mesoscopic devices. Furthermore, the transport behavior predicted by their model is ruled out by the geometry and results of the experiment in Ref. 
\cite{Appugliese2022}. 

\bibliography{biblio}

\appendix
\section{Electrostatics of a GaAs mesa}

\begin{figure}[t]
\centering
\includegraphics[width=\columnwidth]{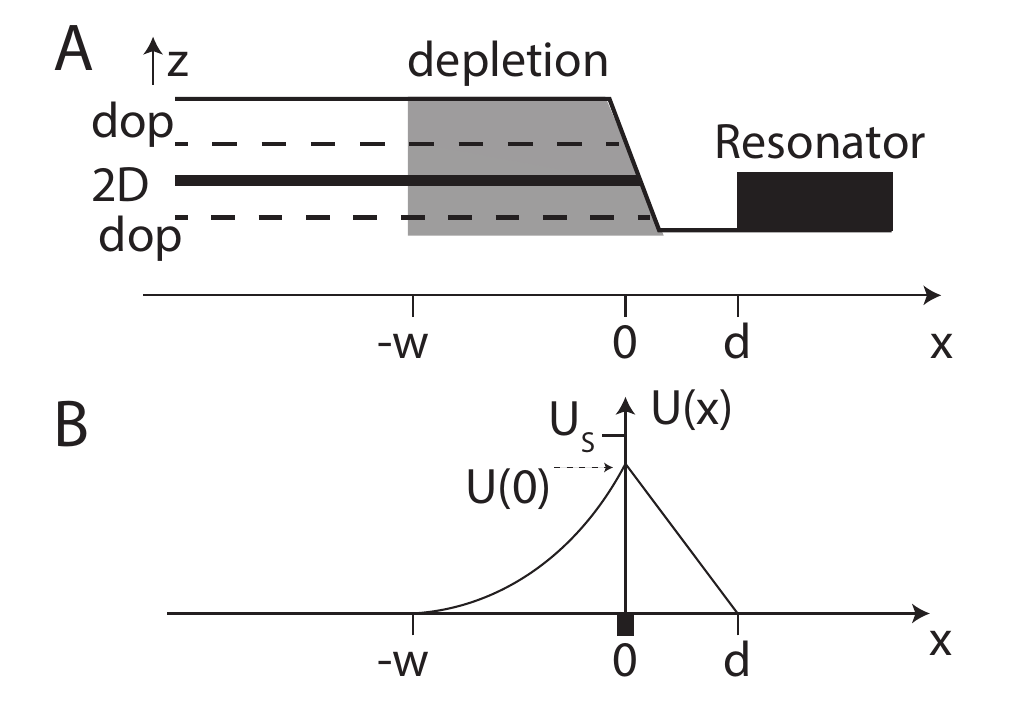}
\caption{
Top: schematic of the device geometry in Ref. \cite{Appugliese2022}.
A semiconductor heterostructure hosts a two-dimensional electron gas (2DEG)
confined in the $x$--$y$ plane, with $z$ the growth direction.
The semiconductor mesa is laterally etched, defining an edge at $x=0$, and is coupled to a nearby
metallic element of a split-ring resonator separated from the semiconductor
surface by an air gap of width $d$.
Ionized donors in the barrier region supply the doping electronic charge.
Bottom: electrostatic potential energy $U(x)$ experienced by an electron
in the 2DEG as a function of the lateral coordinate $x$.
The Schottky barrier $U_S$ is set by the surface states at the GaAs interface.
}
\label{fig:electrostatics-geometry}
\end{figure}

{\it Geometry and electrostatic framework.---}
A sketch of the device is shown in the top panel of
Fig.~\ref{fig:electrostatics-geometry}.
We consider a two-dimensional electron gas (2DEG) lying in the $(x,y)$ plane, with
$z$ denoting the growth direction.
The mesa is laterally etched at $x=0$, the semiconductor occupies the half-space
$x<0$, and translational invariance along the $y$ direction is assumed.
Ionized donors in the barrier supply the electronic charge to the 2DEG.
No external electrostatic bias is applied, and electron depletion occurs in a
region close to the semiconductor--air interface.

The electrostatics of the system is governed by the full two-dimensional Poisson
equation in the $(x,z)$ plane. This is supplemented by the boundary conditions imposed by
Fermi-level pinning at the etched GaAs/air interface due to the presence of a large density of surface states and, when present, by nearby
metallic structures.
Introducing the electron electrostatic potential energy
$\widetilde U(x,z)\equiv -e\,\phi(x,z)$, Poisson’s equation reads

\begin{equation}
(\partial_x^2+\partial_z^2)\widetilde U(x,z)
=
\frac{e^2}{\epsilon_r\epsilon_0}
\left[
N_D^+(z)
-
n_{2D}(x)\,|\chi(z)|^2
\right],
\label{eq:Poisson_2D}
\end{equation}
where $N_D^+(z)$ is the three-dimensional density of ionized donors,
$n_{2D}(x)$ is the two-dimensional electronic density of the 2DEG, and $\chi(z)$ is
the normalized confinement wave function of the lowest quantum well subband.
Here $\epsilon_0$ denotes the vacuum permittivity and $\epsilon_r$ the relative
dielectric constant of the semiconductor.

{\it Effective 1D description with depletion.---}
To provide a simple and illustrative analytical solution, we consider an effective one-dimensional
description within the  depletion approximation, assuming that the electron density is depleted in the interval $-w < x < 0$ where $w$ is depletion length.

In the depletion
region the effective $1D$ Poisson’s equation reads
\begin{equation}
\frac{d^2U(x)}{dx^2}=\frac{e^2 n_{3D}}{\epsilon_r \epsilon_0},
\qquad -w<x<0,
\label{eq:PoissonU}
\end{equation}
where $+e n_{3D}$ is an effective uniform space-charge density which encodes the nonlocal electrostatic effect of remote ionized donors distributed along the growth direction.
The general solution is
$U(x)=\frac{e^2 n_{3D}}{2\epsilon_r\epsilon_0}x^2+Ax+U_0$.
The depletion approximation imposes that the electric field vanishes at the edge of
the depleted region, $U'(-w)=0$, which fixes
$A=e^2 n_{3D}w/(\epsilon_r\epsilon_0)$.
Choosing the bulk reference $U(-w)=0$ then yields
\begin{equation}
U(x)=\frac{e^2 n_{3D}}{2\epsilon_r\epsilon_0}(x+w)^2,
\qquad -w<x<0.
\label{eq:U_semicond}
\end{equation}

{\it Strong pinning boundary condition at the etched GaAs interface.---}
For etched GaAs surfaces, the  high density of mid-gap surface states pins the Fermi
level close to mid-gap~\cite{Heinzel_Mesoscopic_Wiley,WeberPasquarello2014}. This corresponds to a
fixed surface band bending described by a positive Schottky barrier energy $U_S>0$,
such that $U(0)=U_S$ ($\simeq 0.8 eV$ in GaAs). The depletion width then follows as
\begin{equation}
w=\sqrt{\frac{2\epsilon_r\epsilon_0 U_S}{e^2 n_{3D}}}.
\label{eq:w_strong}
\end{equation}

{\it Relaxed pinning and nearby metal.---}
If Fermi-level pinning at the etched GaAs surface is partially relaxed, the surface
charge can vary with the local surface potential energy $U(0)$. This effect is conveniently
parametrized by a finite surface-state capacitance per unit area $C_{ss}$, which
describes the linear response of the surface charge density to variations of the
surface electrostatic potential around the pinned value.  Gauss’s law at the
semiconductor surface yields the boundary condition
\begin{equation}
\epsilon_r\epsilon_0
\left.\frac{dU}{dx}\right|_{0^-}
=
C_{ss}\bigl[U_S-U(0)\bigr]
+ 
\epsilon_0 \bigl[U_m - U(0)\bigr]/d,
\label{eq:BC_U}
\end{equation}
where $U_m=-eV_m$ is the electrostatic potential energy associated with the metal.
The terms on the right-hand side represent, respectively, the contribution of the
surface-state charge and the displacement field $\epsilon_0 \left.\frac{dU}{dx} \right|_{0^+}$across the air gap. The so-called
image-charge effects are nothing but electrostatic consequences of these boundary
conditions and must be treated consistently within Gauss’s law.
Equation~\eqref{eq:BC_U} continuously interpolates between the standard limiting
cases. In the limit of an infinitely large density of surface states,
$D_s(E_F)\to +\infty$, the surface-state capacitance per unit area $C_{ss}=e^2 D_s(E_F)$ diverges and
Eq.~\eqref{eq:BC_U} enforces strong Fermi-level pinning, corresponding to the
boundary condition $U(0) \to U_S$ at the etched GaAs/air interface. Conversely, if the
metallic structure is moved far away, $d\to + \infty$, the geometric capacitance per unit area
$C_{\rm geom} =  \epsilon_0/d$ vanishes and the metal-induced term in
Eq.~\eqref{eq:BC_U} drops out, recovering the boundary condition determined solely
by the GaAs surface states. Note that $U(0) < U_S$ corresponds to a surface with a negatively charged sheet density. 

For an unbiased metal ($U_m=0$), introducing the total capacitance
$C_{\rm tot}\equiv C_{ss}+C_{\rm geom}$, the solution of the depletion width reads
\begin{equation}
w= \sqrt{
\frac{2\epsilon_r\epsilon_0}{e^2 n_{3D}}\,
\frac{C_{ss}}{C_{\rm tot}}\,U_S + 
\left(\frac{\epsilon_r\epsilon_0}{C_{\rm tot}}\right)^2
}- \frac{\epsilon_r\epsilon_0}{C_{\rm tot}} \,.
\label{eq:w_relaxed}
\end{equation}
For etched GaAs surfaces, the density of mid-gap surface states \cite{Heinzel_Mesoscopic_Wiley}
$D_s(E_F)\sim 10^{18}$--$10^{19}\,\mathrm{m^{-2}eV^{-1}}$
corresponds to a surface-state capacitance per unit area
$C_{ss}=e^2 D_s(E_F)\sim 10$--$100\,\mu\mathrm{F\,cm^{-2}}$.
By contrast, for an air gap $d=250\,\mathrm{nm}$,
$C_{\rm geom} \approx \epsilon_0/d\simeq 3.5\,\mathrm{nF\,cm^{-2}}$.
Thus $C_{ss}\gg C_{\rm geom}$ by several orders of magnitude, so that $C_{\rm tot} \simeq C_{ss}$ to excellent accuracy. The metal-induced geometric capacitance $C_{\rm geom}$ therefore
produces only a negligible correction to the depletion width, while no local minimum or attractive pocket is generated by the metal.
{\it Beyond the depletion approximation.---}
Going beyond the depletion approximation requires a fully
self-consistent treatment that couples Poisson’s equation (\ref{eq:Poisson_2D}) for the
electrostatic potential to the equation governing the electronic charge
density.
This can be achieved, for example, within the Thomas--Fermi approximation
or by means of Schr\"odinger--Poisson approaches.
In the presence of Landau quantization, a self-consistent
electrostatic theory have been developed 
for instance in Ref.~\cite{Glazman1992}.

By contrast, the approach of Ref. \cite{Andolina2025} does not constitute a simplified or approximate treatment of the electrostatics of a two-dimensional electron gas. Instead, it replaces the electrostatics of a macroscopic semiconductor system by that of an isolated electron interacting with a nearby metal via an image-charge potential. As a result, it omits at the most basic level the essential ingredients that govern the electrostatics of a GaAs two-dimensional electron gas, namely the boundary conditions imposed by Fermi-level pinning
at the etched GaAs surface,   the positive donor background density and the electron density. 

\end{document}